\documentclass[prb,a4paper,twocolumn,showpacs,floatfix]{revtex4-1}
\usepackage[utf8]{inputenc}
\usepackage[T1]{fontenc}
\usepackage{lmodern}
\usepackage{amssymb}
\usepackage{amsmath}
\usepackage{amsbsy}
\usepackage{bm}
\usepackage{esint}
\usepackage{graphicx}
\graphicspath{ {figs/} }
\usepackage{color}
\newcommand{\I}{\mathrm{i}}
\newcommand{\E}{\mathrm{e}}
\begin{document}
\title{Anomalous quantum Hall effect induced by disorder in topological insulators}
\author{Laurent Raymond}
\author{Alberto D.\ Verga}\email{Alberto.Verga@univ-amu.fr}
\author{Arnaud Demion}
\affiliation{Université d'Aix-Marseille, IM2NP-CNRS, Campus St.\ Jérôme, Case 142, 13397 Marseille, France}
\date{\today}
\begin{abstract}
We investigate a transition between a two-dimensional topological insulator conduction state, characterized by a conductance $G=2$ (in fundamental units $e^2/h$) and a Chern insulator with $G=1$, induced by polarized magnetic impurities. Two kinds of coupling, ferro and antiferromagnetic, are considered with the electron and hole subbands. We demonstrate that for strong disorder, a phase $G=1$ exists even for ferromagnetic order, in contrast with the prediction of the mean field approximation. This result is supported by direct numerical computations using Landauer transport formula, and by analytical calculations of the chemical potential and mass renormalization as a function of the disorder strength, in the self-consistent Born approximation. The transition is related to the suppression of one of the spin conduction channels, for strong enough disorder, by selective spin scattering and localization. 
\end{abstract}
\maketitle

%
\section{Introduction}

Two-dimensional topological insulators are a particular state of matter characterized by the coexistence of insulating bulk states and dissipationless conducting helical edge states.\cite{Cayssol-2013sf,Fruchart-2013yg} This results from the inversion of the valence and conduction bands together with a strong spin-orbit coupling: in helical states the spin and momentum of the carriers are intimately correlated. A typical material is HgTe, which, when confined in a quantum well, exhibits the quantum spin Hall effect.\cite{Bernevig-2006vn,Konig-2007qv} Under a potential bias, two edge states from the two Kramers pairs belonging to a Dirac cone, contribute to the electric current; the existence of a bulk gap ensures the quantization of the conductance to the value of \(G=2\) in units of \(e^2/h\), as can be demonstrated using the Landauer-Büttiker formula.\cite{Shen-2013fk} The quantization of the conductance is a physical phenomenon analogous to the anomalous quantum Hall effect, but in a system invariant under time reversal.\cite{Haldane-1988ys,Kane-2005qf} This quantization is related to the topology of the energy bands; actually, the integer factor \(n\) in \(G=n\, [e^2/h]\) is a Chern number characterizing the total flux of the Bloch wave function (a vector field) over the Brillouin zone (\(n=2\) for the helical edge states of a quantum spin Hall insulator).\cite{Hasan-2010fk}

Topological insulators, in addition to exhibit a wealth of fundamental physical phenomena at the frontier of condensed matter and relativistic field theory,\cite{Qi-2011fk} are a promising material for a variety of applications. In particular, their property of coupling spin and momentum, is ideally suitable for applications in the domain of spintronics, whose goal is to control the spin degree of freedom by pure electrical means. The conduction states of a topological insulator can be used in new concepts of electronic devices, ranging from spin transistors to fast, high density, memories.\cite{Yokoyama-2009ty,Krueckl-2011eu,Mellnik-2014qq,Meng-2014fk} One may also imagine to exploit their ability to support different quantized conduction regimes according to the number of protected edge states, for instance, when doped with magnetic impurities.

Indeed, as a function of the magnetic ordering, a topological insulator doped with a transition metal can support a quantum anomalous Hall state, phase similar to the quantum Hall state but without an external magnetic field.\cite{Liu-2008ys,Yu-2010kx,He-2014bh} Therefore, the system can, in principle, change from a \(G=2\) state in its topological time reversal symmetric phase (``Dirac'' spin quantum Hall state), to a \(G=1\) state in an anomalous quantum Hall phase (Chern insulator),  to eventually a normal insulating or metallic state. The observation of an anomalous quantum Hall state in a thin film of a topological insulator was recently realized experimentally.\cite{Chang-2013ys} 

In this paper, our objective is to demonstrate that a two-dimensional topological insulator doped with magnetic impurities undergoes a disorder driven quantum phase transition between these topologically different states, when the disorder strength is varied. We start by studying a model of a HgTe quantum well, with uniformly distributed magnetic moments polarized perpendicularly to the plane. Numerical transport calculations were performed using the Landauer formula. We investigate the conductance in a two terminals setup, as a function of the disorder strength and the chemical potential. We focused in particular on the behavior of the edge channels according to their spin. Finally, we calculate, in the second order self-consistent Born approximation, the real part of the self-energy to obtain explicit expressions of the renormalized mass and the chemical potential.

\section{Model}

A simple model of a two-dimensional topological insulator is given by a four band Hamiltonian of the form (in momentum space),\cite{Bernevig-2006vn}
\begin{equation}
  \label{e:TBH0}
  H_0(\bm k)=\begin{pmatrix}
    h_\uparrow(\bm k) &    0     \\
    0    &  h_\downarrow(\bm k)  
\end{pmatrix}\,,
\end{equation}
where
\begin{multline}
  \label{e:TBh}
  h_\uparrow(\bm k)= \frac{A}{a} (\sin (ak_x) \tau_x + \sin(a k_y) \tau_y) 
  + m\tau_z \\ 
  - \frac{2}{a^2}(D\tau_0 + B\tau_z) [2-\cos(a k_x) - \cos(a k_y)] \,,
\end{multline}
and \(h_\downarrow(\bm k) = h_\uparrow(-\bm k)^*\) in the (spin\(\otimes\)band)\(=\bm \sigma \otimes \bm \tau\) base, \(\bm \tau=(\tau_x, \tau_y, \tau_z)\) and \(\bm \sigma=(\sigma_x,\sigma_y,\sigma_z)\) are Pauli matrices in band and spin spaces (with \(\tau_0,\sigma_0\) identity matrices, and the Kronecker product is denoted by a pair as in \(\sigma_x \tau_y\)), \(\bm k = (k_x,k_y)\), \(k=|\bm k|\) is the wavenumber, \(a\) the lattice step, and \(A,m,B,D\) are material parameters. In the following we adopt units such that \(a=A=\hbar=1\). In such a system, typical values are \(m = -0.137\), \(B=-0.376\) and \(D=-0.281\), as computed with \(a=5\,\mathrm{nm}\) and \(A=364\,\mathrm{nm}\,\mathrm{meV}\), which are standard tight-binding parameters of HgTe thin films (the unit of energy is \(\varepsilon_0=73\,\mathrm{meV}\)).\cite{Konig-2008xy} The form of the Hamiltonian insures invariance with respect to time reversal and an inversed band structure when \(m<0\) (\(mB>0\)).

In order to take into account the disorder, we add an exchange term \(J_I\) coupling to the impurity's normalized magnetic moment \(\bm S_i\), \(S_i=1\), at lattice site \(i\) (in position representation):
\begin{equation}
  \label{e:vx}
  V_\alpha = J_I \sum_{i \in I} 
  \bm c^\dagger_i (\bm S_i \cdot \bm \sigma\,\tau_\alpha) c_i
\end{equation}
where \(c_i=(c_{i+\uparrow}, c_{i-\uparrow},c_{i+\downarrow}, c_{i-\downarrow})\) is the annihilation operator at position \(\bm x_i\) of ``electrons'' \((+)\) and ``holes'' \((-)\) with spin components up \((\uparrow)\) and down \((\downarrow)\). The sum is over the set of \(N_I\) impurity sites \(I\) uniformly distributed in the lattice; we denote \(n_I=N_I/N\) their concentration (\(N\) is the number of lattice sites). The parameter \(\alpha = 0, z\), determines the type of magnetic coupling with the band states: for \(\alpha=z\) the spin splitting is of opposite sign for electrons and holes, we shall refer to this case as ``antiferromagnetic,'' and for \(\alpha=0\), both quasiparticles have the same Zeeman splitting, this case shall be referred as ``ferromagnetic.''\cite{Liu-2008ys} The impurity magnetic moment is randomly oriented inside a cone of angle \(\theta_0\) around the \(z\)-axis:
\[
  \langle \bm S \rangle = (0,0,M_z)\,,
  \langle S_z^2 \rangle = \frac{1}{3} \left[2M_z(2M_z-1) + 1\right]  \,,
\]
where \(M_z=\cos^2(\theta_0/2)\) is the mean magnetization, and we have taken into account that \(\bm S_i\) is modulus one. Imposing a magnetic order breaks the time reversal symmetry, modifying the electronic properties of the edge states. In a mean field approximation, in the ferromagnetic case, the \(\tau_0\) term splits the edge states but do not open a gap, while in the antiferromagnetic case the term in \(\tau_z\) opens in addition a gap (for one of the two spin polarizations). One may infer that the former case is trivial and the later one transform the topological insulator into a Chern insulator.\cite{Liu-2008ys} This picture can be deeply modified by spin dependent backscattering and by localization effects due to disorder.

\begin{figure*}
  \centering
  \includegraphics[width=0.49\textwidth]{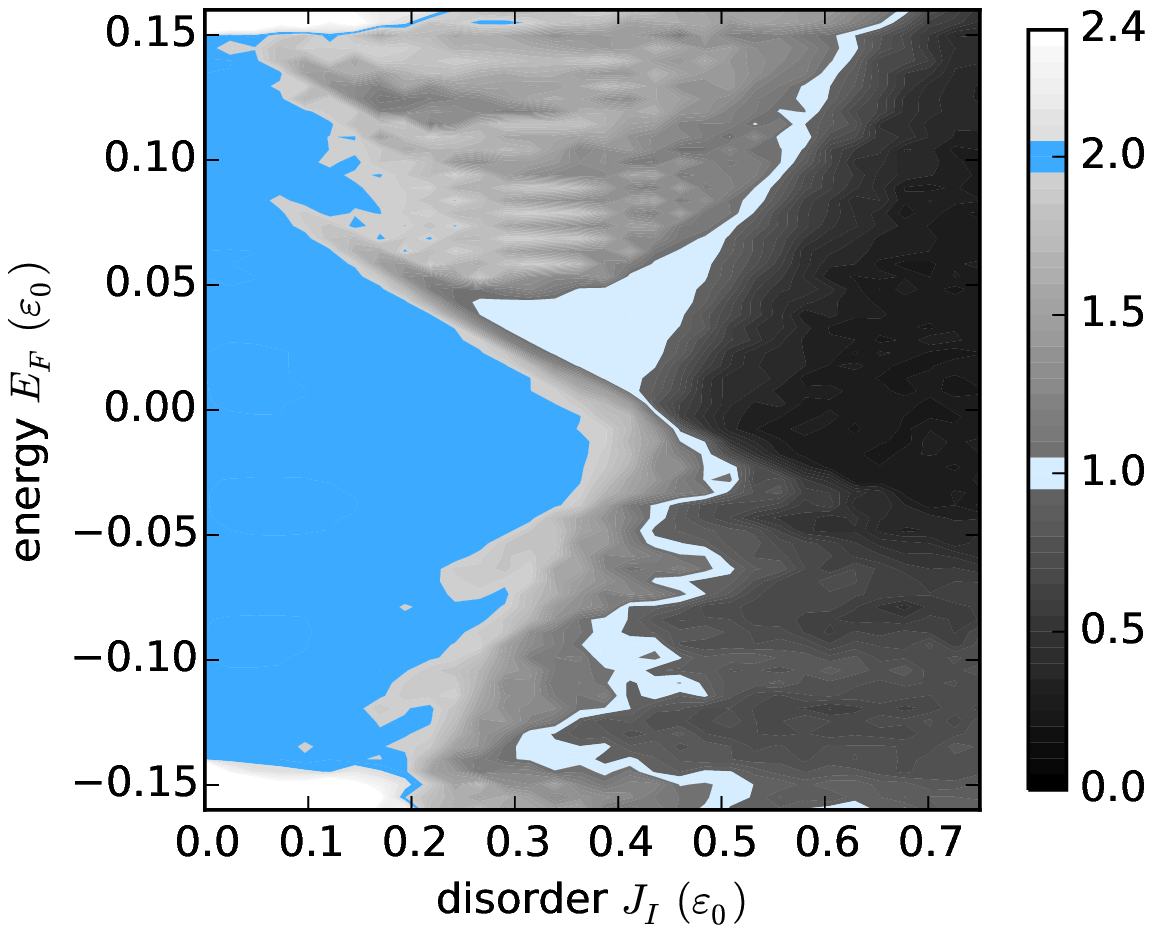}\hfill%
  \includegraphics[width=0.49\textwidth]{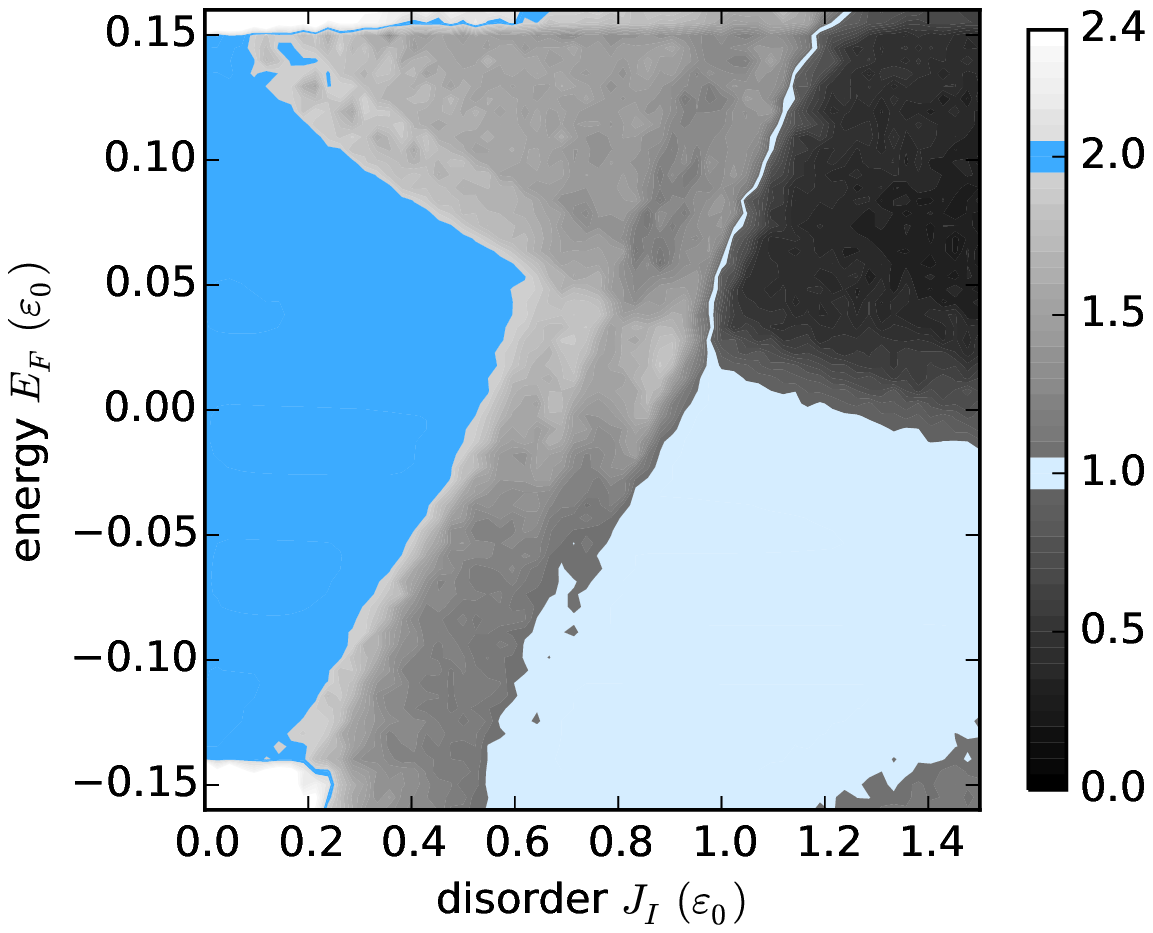}
  \caption{\label{f:ti1}\small Conductance phase diagram in the Fermi energy \(E_F\), disorder strength \(J_I\) parameter space, for the antiferromagnetic (left) and ferromagnetic (right) subband couplings. The topological insulator phase, for \(E_F \approx 0\) in the bulk gap ($|E_F| < |m|=0.137$), is characterized by \(G=2\) (in dark blue) at weak disorder; a Chern insulator state \(G=1\) (in light blue) appears at stronger noise intensity in both cases. The polarization of impurities is \(M_z=1\), and their concentration is \(n_I=0.4\) for the antiferromagnetic case, and \(n_I=0.2\) for the ferromagneitic case.}
\end{figure*}

\section{Results}

Indeed, our numerical computations of the tight-binding model (\ref{e:TBH0}-\ref{e:vx}), demonstrate that in both cases, there is a transition from Dirac to Chern states. 

We computed the conductance as a function of the Fermi energy \(E_F\) and the disorder strength \(J_I\) from the Landauer-Büttiker formula using nonequilibrium Keldysh Green functions and a recursive method,\cite{Datta-1997fk,Nikolic-2009kx} The conductance \(G\), is computed in the linear response approximation, from the retarded \(G^R\) and advanced \(G^A\) Green functions,\cite{Caroli-1971xp,Meir-1992vn}
\[
  G = \frac{e^2}{h} \int_{-\infty}^{\infty} dE \frac{\partial f}{\partial E } 
  \big\langle
   \mathrm{Tr}\big[ \Gamma_R G^R(E) \Gamma_L G^A(E) \big]
  \big\rangle\,, 
\]
where \(f=f(E)\) is the Fermi-Dirac distribution, \(\Gamma_{L,R}\) are the broadenings due to the left ($L$) and right ($R$) leads, the trace is taken over the band and spin indices, and the angle brackets are for the disorder averaging.  Our code allows also the computation of local quantities, such as the density of quasiparticles, the density of states and the currents. Specifically, we calculated the transport through a disordered central region, connected to clean topological insulator leads. The lattice size is \(72+256+72\times64\), and the physical quantities were averaged over a set of different impurities distributions. To minimize the effects of discontinuities, an intermediate clean region was inserted between the semi infinite leads and the doped region. We explored a range of Fermi energies around the gap of the clean system, and exchange coupling strengths up to the strong disorder regime (Fig.~\ref{f:ti1}).

If one replaces the random potential (\ref{e:vx}) by its mean value (proportional to the polarization \(M_z\)), the resulting Hamiltonian can be readily diagonalized, and the modification of the energy bands predicted. When the coupling of electrons and holes are of different signs (subband antiferromagnetic coupling), the spin up band develops a gap at a critical value of the disorder (in our case, \(J_I n_I M_z=0.137=-m\)). In the other case (subband ferromagnetic coupling), bands of opposite spin cross without opening a gap. Therefore, the mean field approximation predicts the disappearance of one spin channel in the antiferromagnetic case, and a simple transition to a normal state in the ferromagnetic case. In both cases, the topological insulator bulk gap vanishes linearly with the noise strength (the chemical potential is a linear function of \(J_I\)). 

The numerical computations represented in Fig~\ref{f:ti1}, qualitatively confirm this scenario. We show the conductance in the parameter range corresponding to the  bulk gap (\(|E_F| < |m| = 0.137\)) of the clean system, from weak to strong disorder, and for the two kinds of magnetic coupling. In particular, the antiferromagnetic case (left panel), shows regions in the \((J_I,E_F)\) plane, with \(G=2\) (dark blue), and \(G=1\) (light blue), corresponding to transport with two active channels (topological insulator state) and one active channel (Chern insulator state), respectively.  However, in the ferromagnetic case (right panel), a region with \(G=1\) is observed, in contradiction with the mean field prediction. (Note that the relevant states for the quantized conductance are in the bulk gap, which may disappear for strong polarized disorder.) 

The phase diagram of the antiferromagnetic case shows an anomalous quantum Hall state for disorder strengths in the range \(J_I \approx 0.3\dots0.4\); the mean field prediction is \(J_I \approx 0.35\) (we used \(n_I=0.4\) and \(M_z=1\)). It is interesting to observe that for a fixed value of \(J_I\), it is possible to change the conduction state from \(G=2\) to \(G=1\) by increasing the Fermi energy. In the ferromagnetic case the extension of the \(G=2\) phase is similarly well described by the mean field approximation (with \(n_I=0.2\) and \(M_z=1\) the mean field closing gap value is \(J_I \approx 0.7\)). A large \(G=1\) region appears at strong disorder and mostly negative Fermi energies. The extension and even the existence of these quantized conductance states are obviously dependent on the underlying symmetries of the system. Decreasing the polarization of the impurities (\(M_z<1\)), that is allowing in-plane fluctuations of their magnetic moments, breaks the spin orientation conservation (the Hamiltonian no longer commutes with the vertical component of the spin), and backscattering and spin flipping of the edge states shrinks the topological insulator phase. Simulations for various polarization states (not shown), suggest that the Chern insulator phase is more robust in the case of ferromagnetic subband coupling than in the antiferromagnetic case. As a matter of fact, the response of the system to the random perturbations produced by the spatial inhomogeneities in the distribution of impurities, is qualitatively different in the two cases. The Chern insulator phase for ferromagnetic coupling, being an effect of strong disorder, is naturally insensitive to these fluctuations. In the antiferromagnetic case, we note important fluctuations of the conductivity in the region around the \(G=1\) phase. As a result, the \(G=1\) region itself slightly moves in the \((J_I,E_F)\) space for each noise distribution. Therefore, the effect of averaging is to blur the contours of the Chern phase region, reducing its extension. We verified that fluctuations are completely suppressed in both \(G=2\) and \(G=1\) phases.

\begin{figure*}
  \centering
  \includegraphics[width=0.49\textwidth]{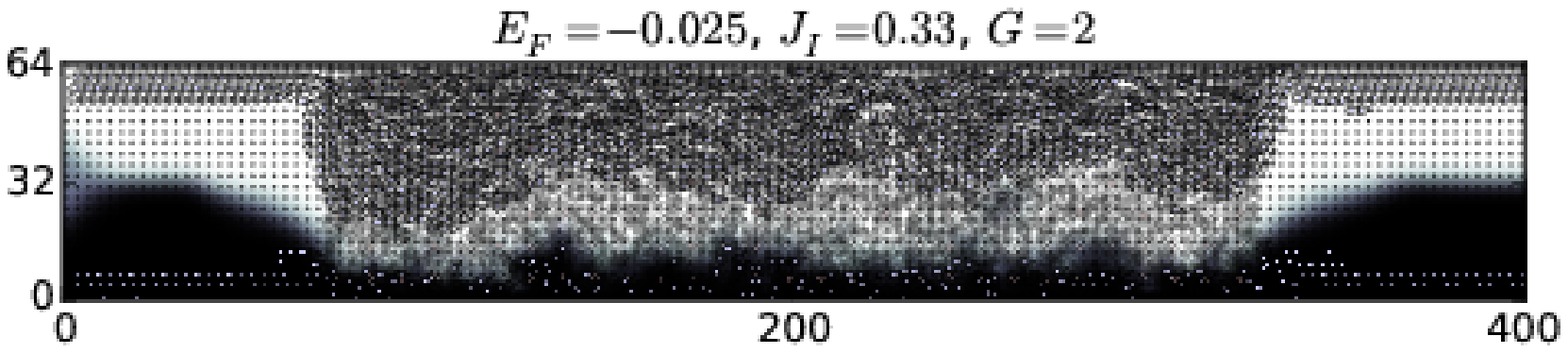}\hfill%
  \includegraphics[width=0.49\textwidth]{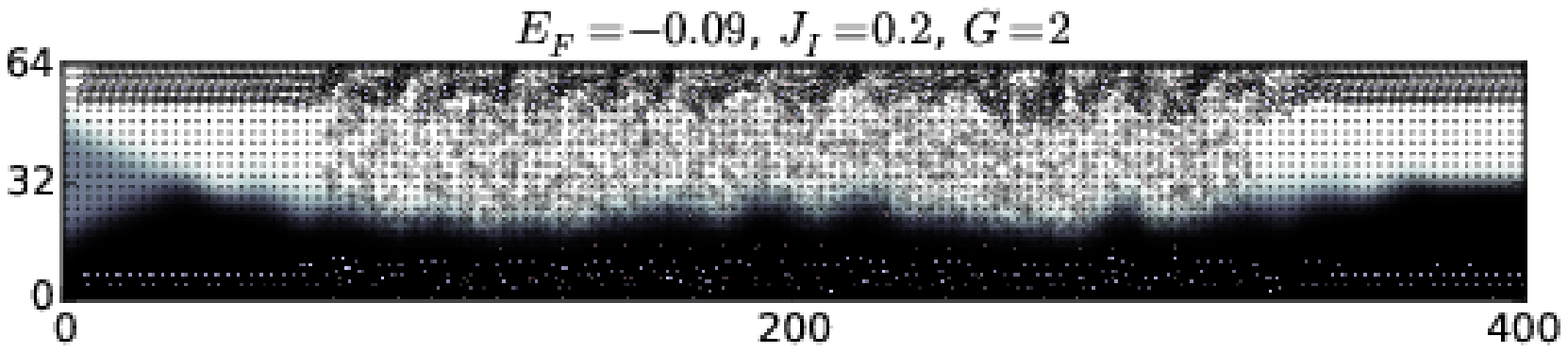}\\
  \includegraphics[width=0.49\textwidth]{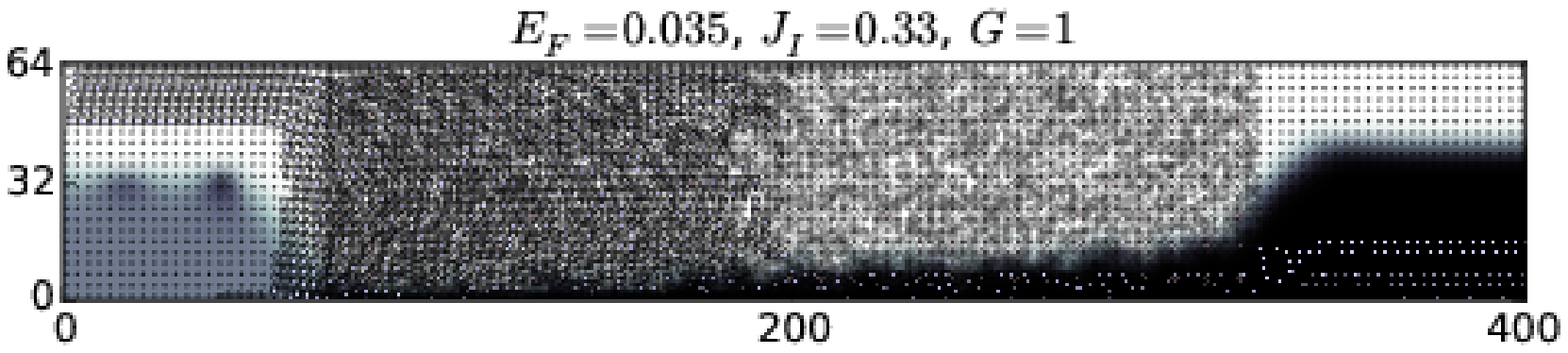}\hfill%
  \includegraphics[width=0.49\textwidth]{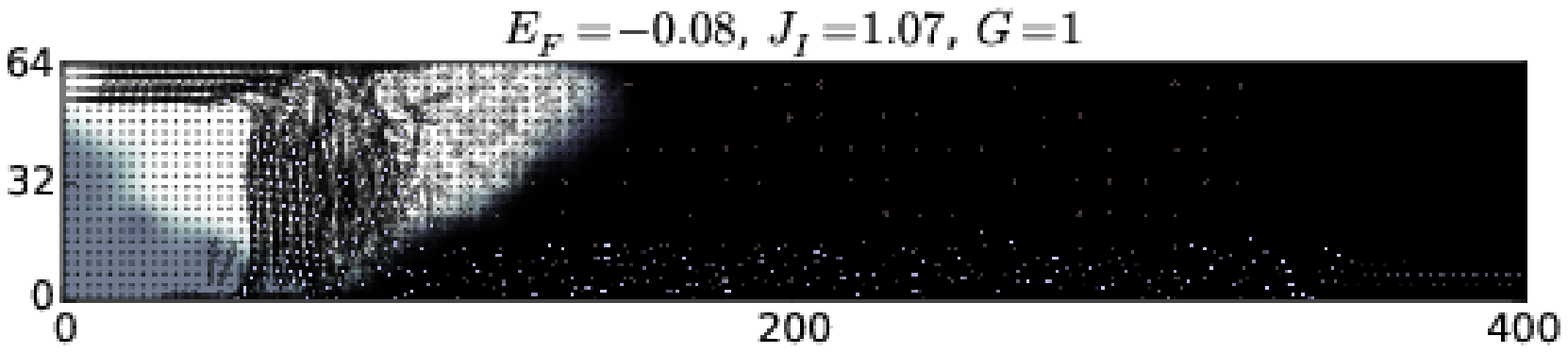}\\
  \includegraphics[width=0.49\textwidth]{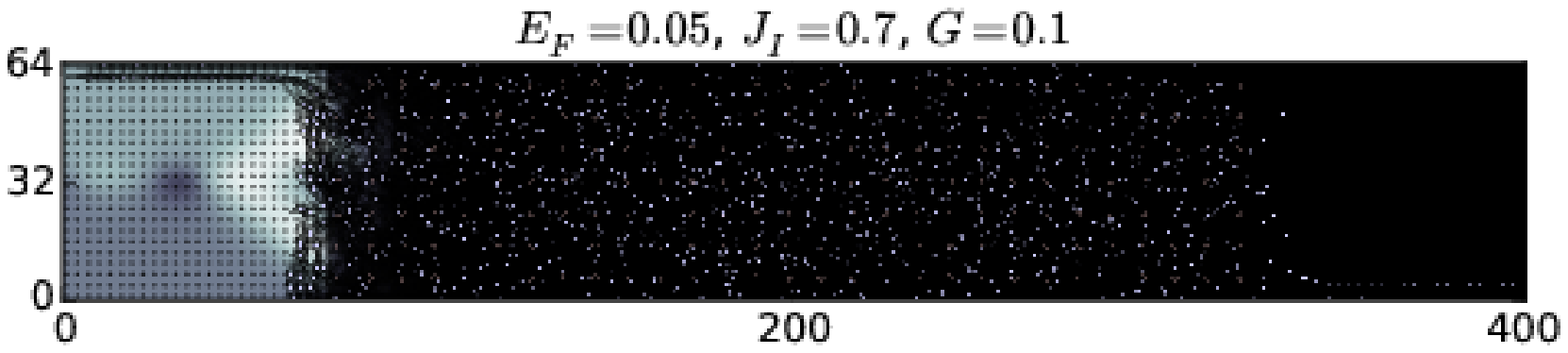}\hfill%
  \includegraphics[width=0.49\textwidth]{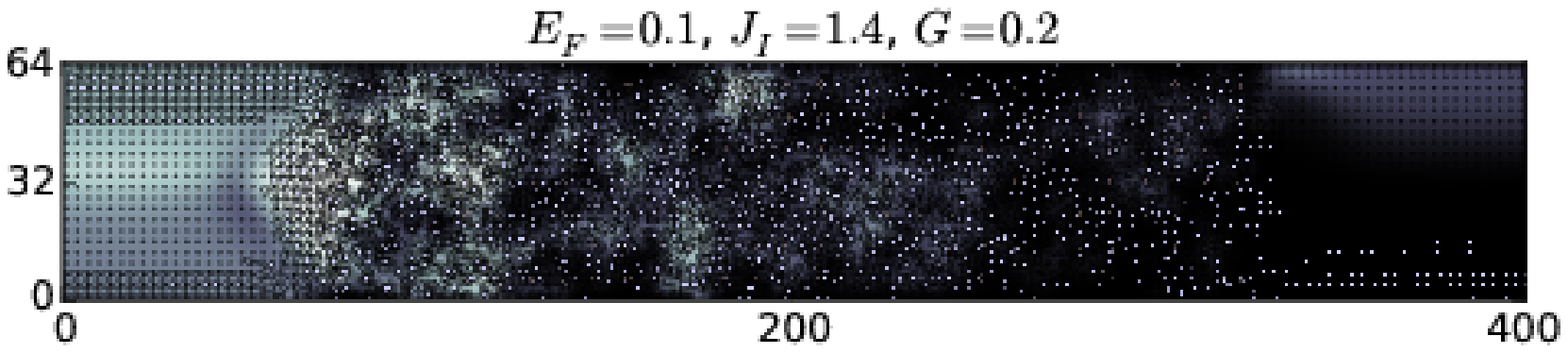}\\
  \includegraphics[width=0.32\textwidth]{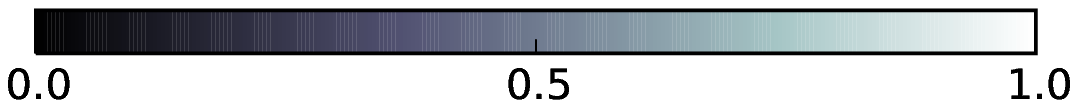}
  \caption{\label{f:ti2} Local currents in various conduction states for the antiferromagnetic (left) and ferromagnetic (right) cases: \(G=2\) (top), \(G=1\) (middle) and \(G \ll 1\) (bottom). The Chern insulating state \(G=1\), exhibits the persistence of spin down edge state. The background color represents the spin-up nonequilibrium density excess $d_\uparrow$, from full spin-up polarization in white, to full spin-down polarization in black.}
\end{figure*}

For fully polarized impurities, the microscopic mechanism of the transitions between different conductance regimes, can be investigated by computing the bond current field,\cite{Caroli-1971xp}
\[
  I_{ij} = \frac{-2e}{\hbar} \int_{-\infty}^{\infty} \frac{dE}{2\pi}
  \mathrm{Tr}[t_{ij}G^<(E,j,i) - t_{ji} G^<(E,i,j)] \,, 
\]
where \(I_{ij}\) is the current between neighboring sites \((i,j)\), \(G^<\) is the lesser Green function, it is a matrix in band and spin indices, and \(t_{ij}\) are the jump matrices, and the nonequilibrium spin density,
\[
  \Delta n_{\uparrow \downarrow}(i) = \int_{\mu-eV}^{\mu+eV} \frac{dE}{2\pi \I}  
    G^<_{\uparrow \downarrow}(E,i,i) \,,
\]
where $V$ is the external bias voltage. The local currents and the spin up density excess 
\[
  d_\uparrow = \frac{ \Delta n_\uparrow}{\Delta n_\uparrow + \Delta n_\downarrow}\,,
\]
are represented in Fig.~\ref{f:ti2} (arrows field and background gray levels, respectively).

The dependence on the relative orientation of the carriers spin and impurities magnetic moment, leads to a rich variety of interactions affecting differently the edge states according to their polarization. In particular, the effect of magnetic disorder is to selectively localize bulk states depending on their spin. Comparing the paths of the carriers through the scattering off impurities region between the antiferromagnetic (left) and ferromagnetic (right) subband coupling cases, we observe that their are qualitatively similar (we used the same numerical parameters in Figs~\ref{f:ti1} and \ref{f:ti2}). The main difference is in the width of the edge channels, especially in the bulk region. In the topological insulator phase (top row) both spin-up top channel and spin-down bottom channel pass through the disordered region; in the antiferromagnetic case the top channel deeply penetrates in the bulk. In the Chern insulator phase (middle row) the suppression of the spin-up component is much stronger in the ferromagnetic case. The non-quantized conductance state is shown in the bottom row. The strong scattering allows the connexion between the two edges, leading to a situation where one of the two spins polarizations is completely filtered out; an accumulation of the other spin polarization can therefore appear on the opposite lead side. The system acts as a spin selection filter in the two cases shown in Fig.~\ref{f:ti2}, \(G=1\) for ferromagnetic symmetry (middle row, left) and \(G<1\) for the antiferromagnetic symmetry (bottom, right).

\begin{figure*}
  \centering
  \includegraphics[width=0.49\textwidth]{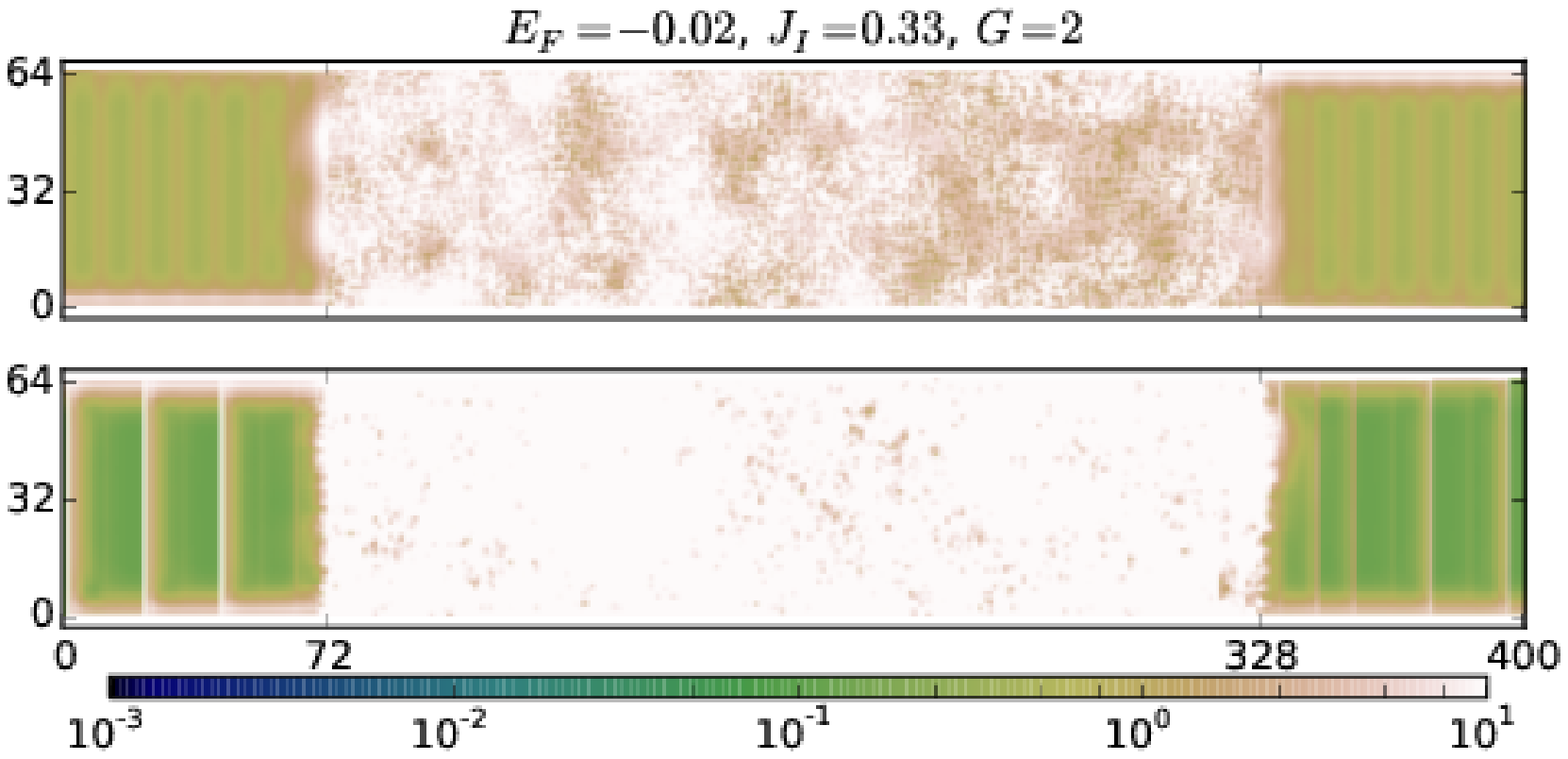}\hfill%
  \includegraphics[width=0.49\textwidth]{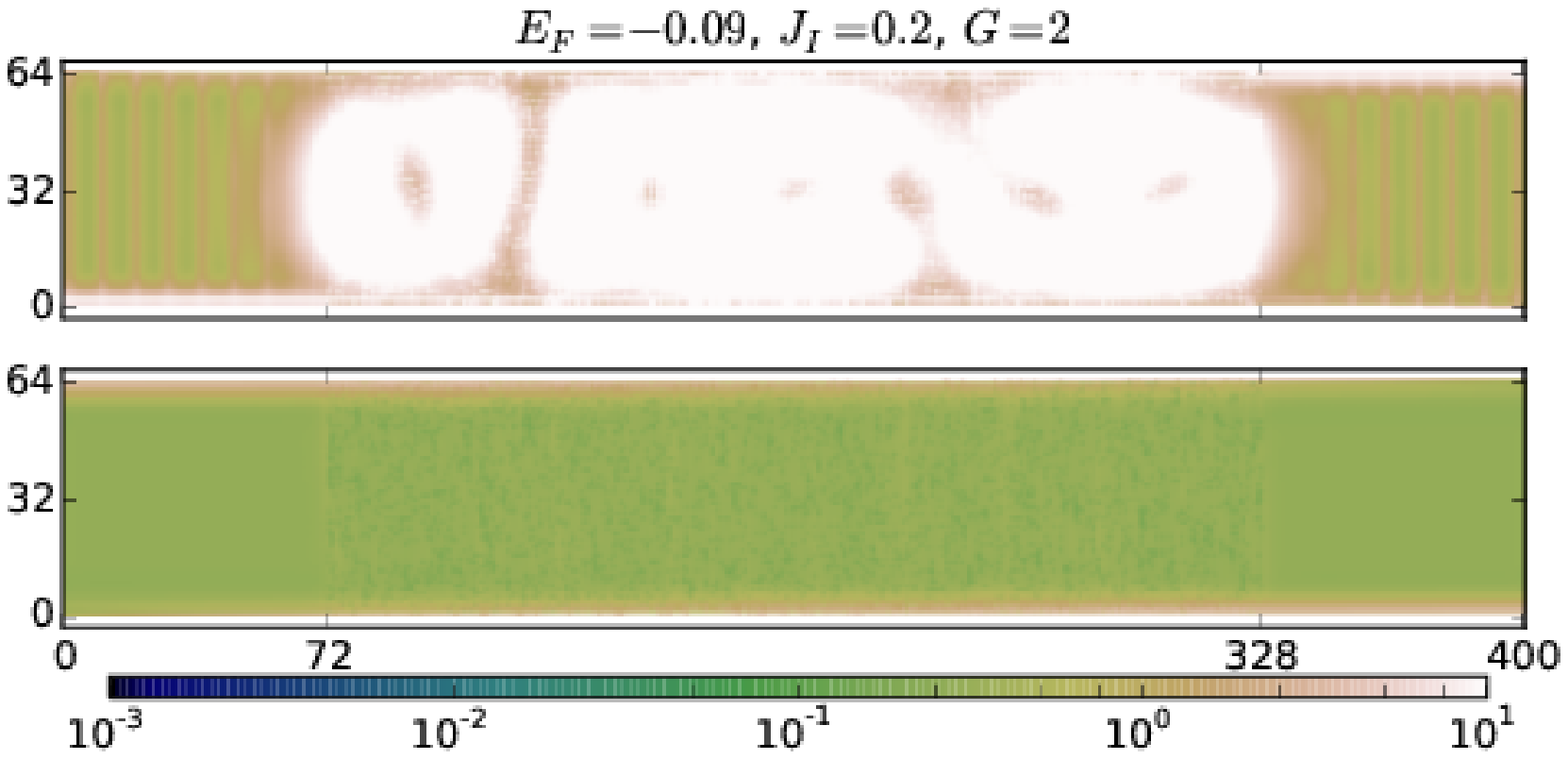}\\
  \includegraphics[width=0.49\textwidth]{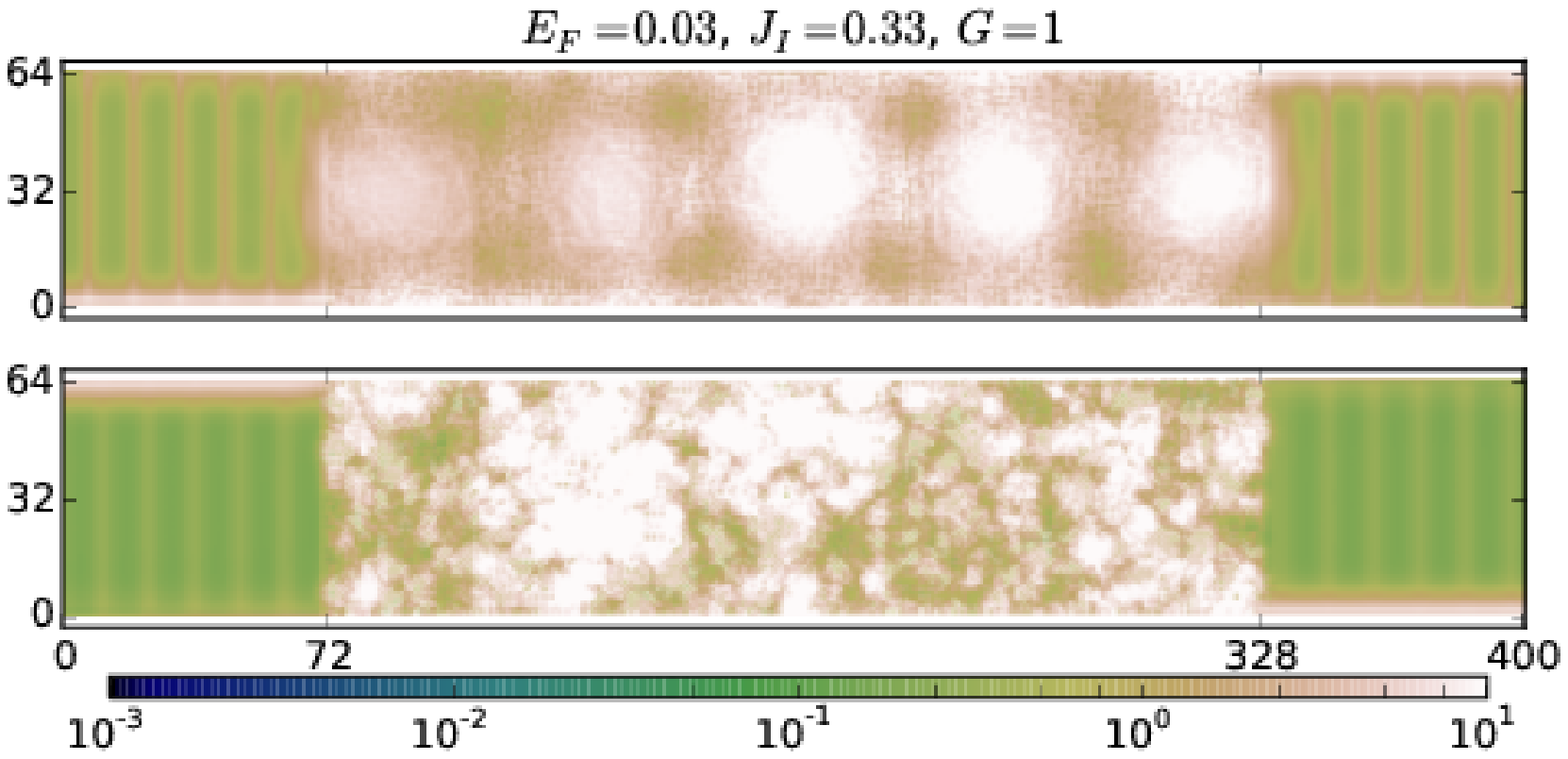}\hfill%
  \includegraphics[width=0.49\textwidth]{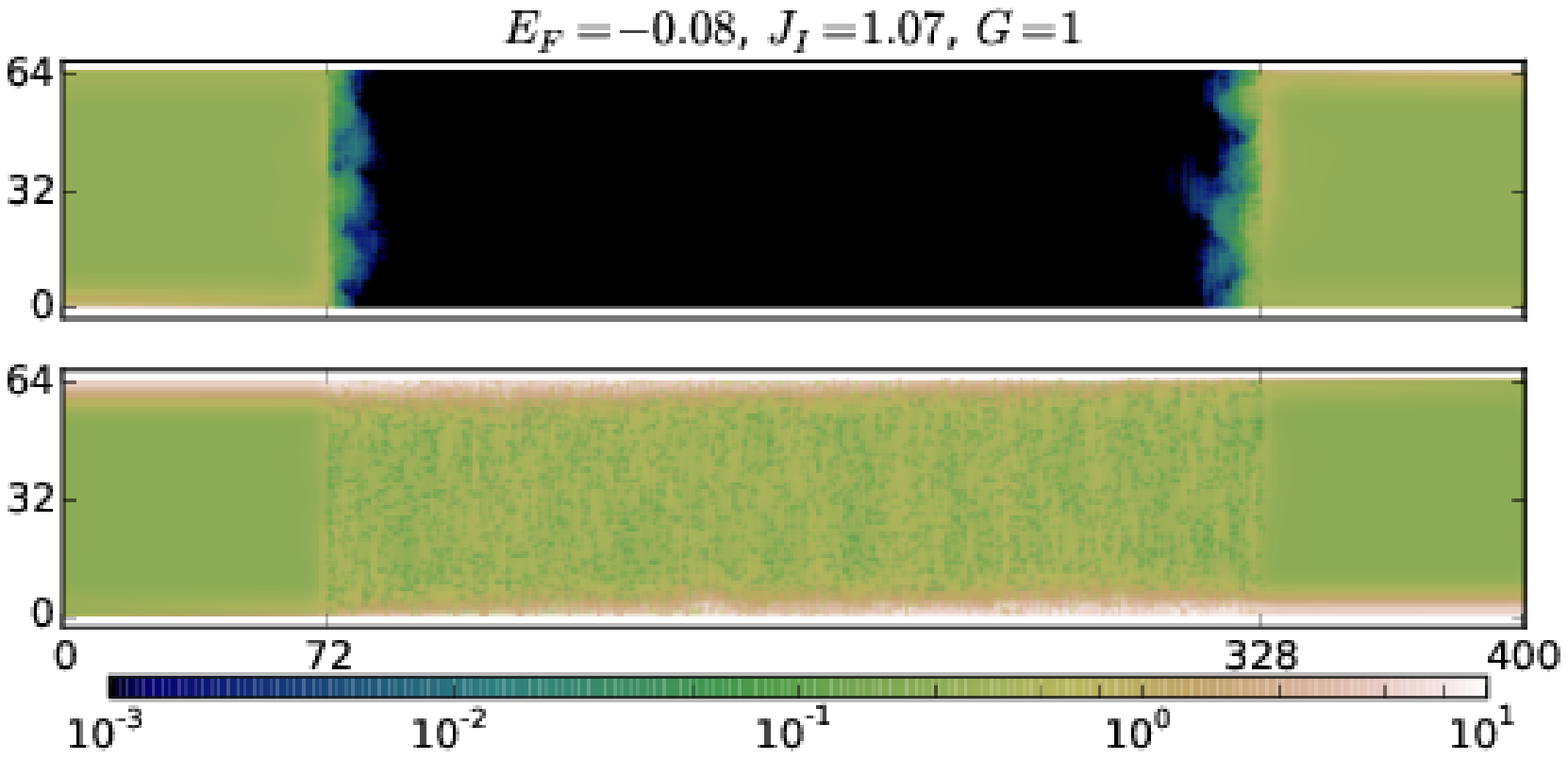}\\
  \includegraphics[width=0.49\textwidth]{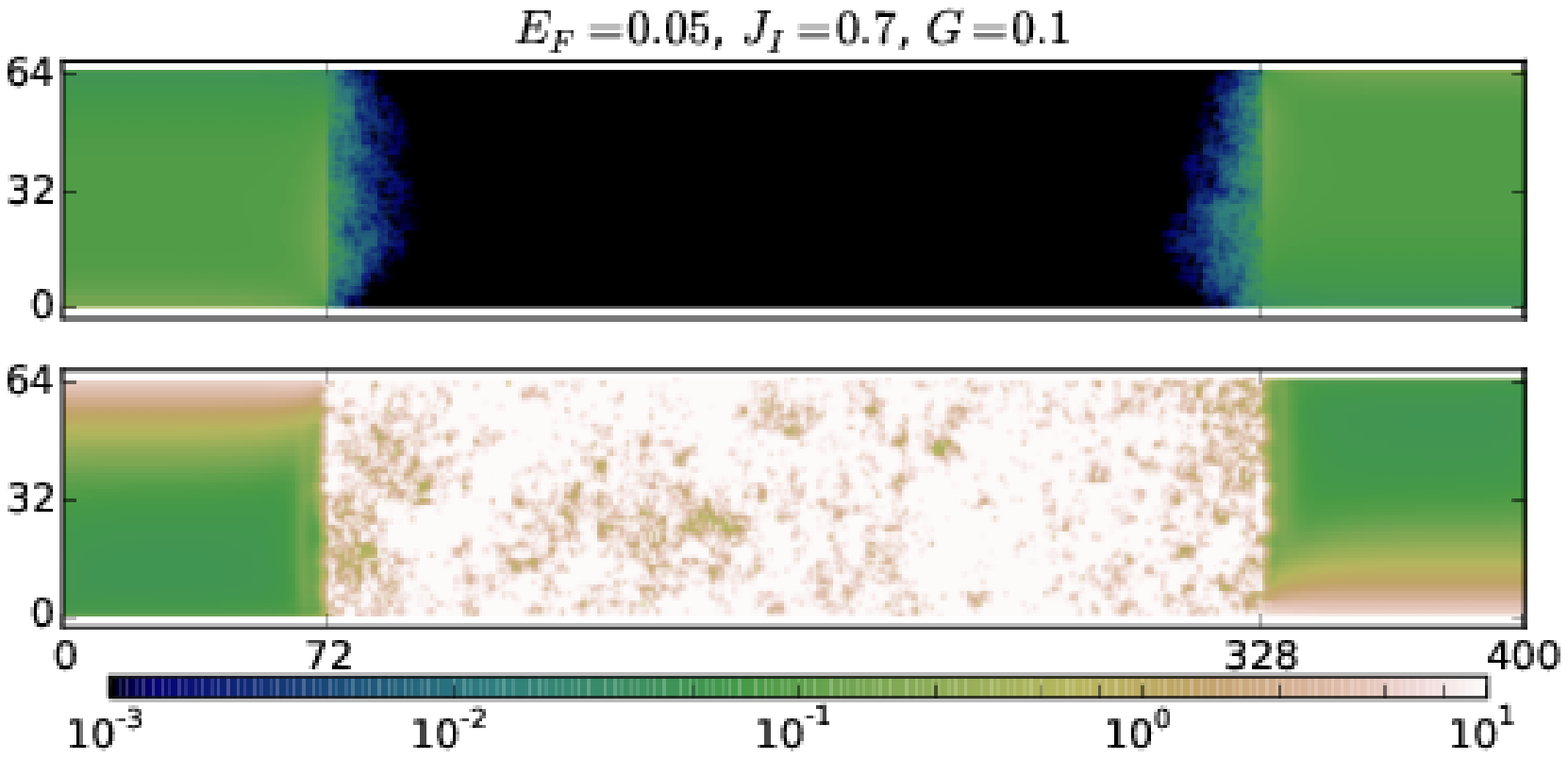}\hfill%
  \includegraphics[width=0.49\textwidth]{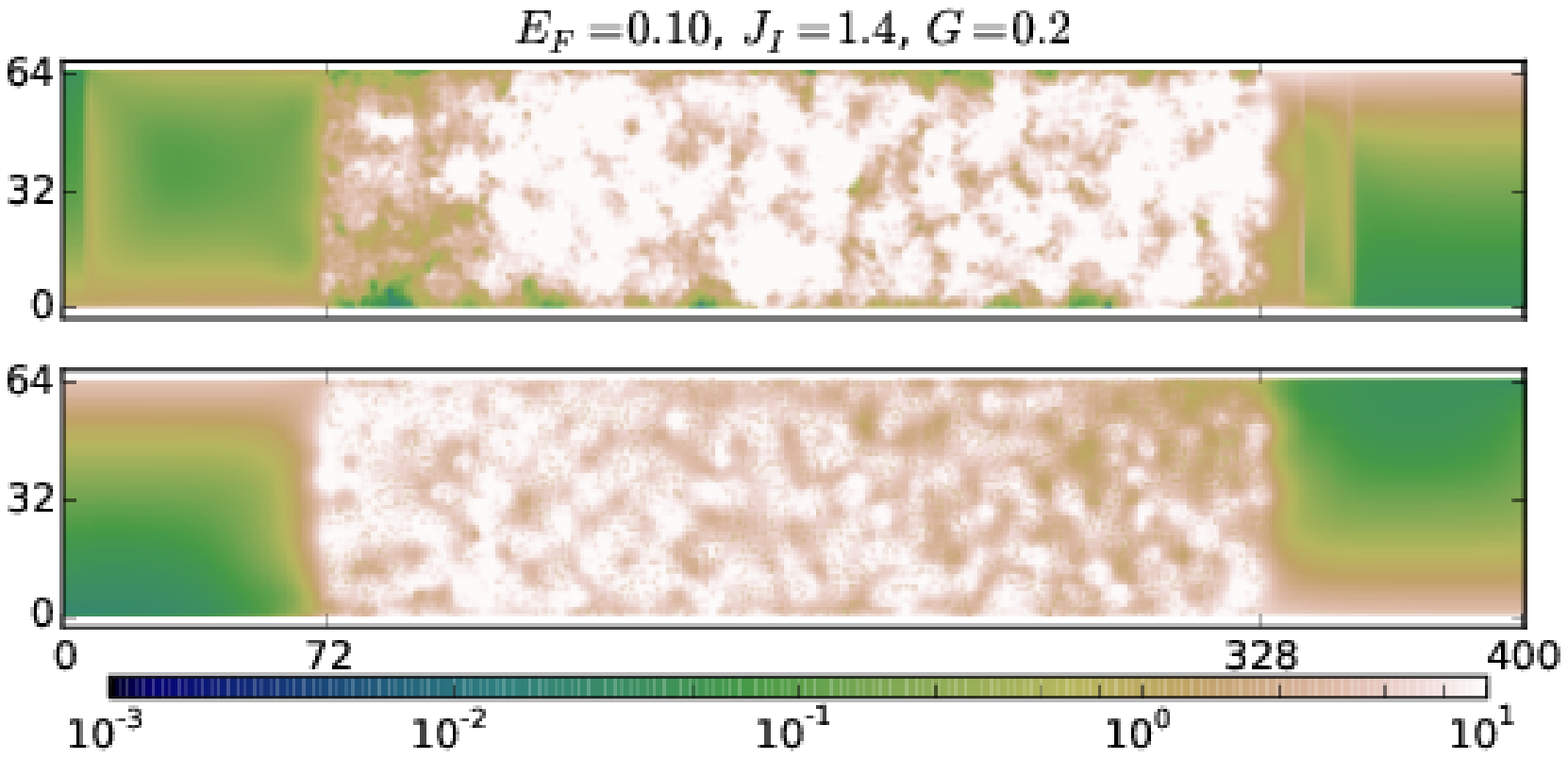}
  \caption{\label{f:ti3} Local density of states associated to the local currents of Fig.~\protect\ref{f:ti2}. Antiferromagnetic (left column) and ferromagnetic (right column) cases. For each transport regime, $G=2$, $G=1$ and $G<1$ top, middle and bottom rows, respectively, the spin-up (top) and spin-down (bottom) components are presented. }
\end{figure*}

These results suggest that the setting up of the anomalous quantum Hall state for antiferromagnetic and ferromagnetic couplings are not due to the same microscopic mechanism. In the antiferromagnetic case it is a consequence of the opening of a band gap for one spin species. The ferromagnetic case is at variance, an effect of the disorder and the establishment of a spin dependent mobility gap. The emergence of localized states can be uncovered by the measure of the local density of states,
\begin{align*}
  \rho_{\uparrow,\downarrow}(E,i) &= \sum_{n} 
    \langle n|c_{\uparrow,\downarrow}(i)^\dag c_{\uparrow,\downarrow}(i)
    |n \rangle \delta(E-\varepsilon_n)\\
    &=
    -\frac{1}{\pi} \mathrm{Im}\, G^R_{\uparrow,\downarrow}(E,i,i)
  \end{align*}
  where \(|n \rangle\) and \(\varepsilon_n\) are eigenstates and eigenvalues of \(H\), and \(G^R(E,i,i)\) the retarded Green function computed at site \(i\). In Fig.~\ref{f:ti3} we show \(\rho_{\uparrow,\downarrow}(E,i)\) in a logarithmic scale, for the same parameters (energy and disorder strength) used in Fig.~\ref{f:ti2}: antiferromagnetic (left) and ferromagnetic (right) cases, for three values of the conductance, \(G=2\) (top), \(G=1\) (middle), and \(G<1\) (bottom). Each panel includes the spin-up (\(\rho_\uparrow\), top) and spin-down (\(\rho_\downarrow\), bottom) components. The presence of open channels are visible in each case: two spins in the $G=2$ conduction regime, and one spin in $G=1$ and normal conduction regimes. The main physical difference between the two coupling modes, antiferromagnetic and ferromagnetic, appears in the \(G=1\) anomalous quantum Hall state. In the antiferromagnetic case, there are energy states available for the two spins polarizations. This is in contrast with the complete absence of available spin-up states in the ferromagnetic case. Therefore, we have evidence showing that the mechanism allowing the anomalous quantum Hall effect for ferromagnetic coupling, do not need a bulk gap but rather a mobility gap and the selective localization of one spin polarization states (in analogy with the topological Anderson insulators\cite{Zhang-2012fk}). For the antiferromagnetic case, localization appears at stronger disorder (see bottom row, left of Fig.~\ref{f:ti2}).

\section{Discussion and conclusions}

An analytical computation of the renormalized Fermi energy and mass as a function of the disorder strength, in the self-consistent Born approximation, allows us to get some insight into the effects of the magnetic impurities on the transport properties and to confirm the qualitative behavior depicted by the numerical simulations. We are specifically interested in the dependence of these renormalized quantities on the spin of the current carriers and their coupling with the spatially distributed magnetic moments. We follow a method used to study the transition of an insulator to a topological insulator, triggered by Anderson localization.\cite{Li-2009kx,Groth-2009vn,Jiang-2009fk}

\subsection{Born approximation}

The thin film Hamiltonian becomes, in the small wavenumber approximation,
\begin{equation}
  \label{e:H4cc}
  H=\sum_{\bm k} c_{\bm k}^\dagger [H_0(\bm k) - \mu]c_{\bm k} +
      \sum_{\bm k, \bm q}  c_{\bm q}^\dagger V_\alpha(\bm k)c_{\bm k+\bm q}\,,
\end{equation}
where we added a chemical potential term \(\mu\) to the unperturbed Hamiltonian \(H_0\). The diagonal blocks of \(H_0\), \(h_{\uparrow,\downarrow}\) of Eq.~(\ref{e:TBH0}), are now given by:
\begin{align}
  h_\uparrow(\bm k) &= -Dk^2\tau_0 + (m-Bk^2)\tau_z + \bm k \cdot \bm \tau 
    \nonumber \\
  h_\downarrow(\bm k) &= -Dk^2\tau_0 + (m-Bk^2)\tau_z - \bm k \cdot \bm \tau^*\,,
  \label{e:hud}
\end{align}
and the random potential is, in Fourier space,
\begin{equation}
  \label{e:Vk}
  V_\alpha(\bm k) = J_I\sum_{i \in I} 
      \bm S_i\cdot\bm \sigma\tau_\alpha 
      \E^{\I \bm k \cdot \bm x_i}  \,.
\end{equation}
We are interested in the renormalization of the mass \(m\) and the chemical potential \(\mu\), which depend on the real part of the self-energy \(\Sigma(E)\),\cite{Groth-2009vn} and eventually in their splitting depending on spin. It is convenient to take into account the mean value of the random potential \(\langle V \rangle\) in the free Green function (the angle brackets stand for the \(\bm S_i\) and \(\bm x_i\) probability distribution integration), 
\begin{equation}
  \label{e:gEk}
  g_\alpha(E,\bm k)= \left[
    E + \mu - H_0(\bm k) - \langle V_\alpha \rangle 
    \right]^{-1}\,,
\end{equation}
where \(\langle V_\alpha \rangle=\langle V_\alpha(0) \rangle\),
\[
  \langle V_\alpha \rangle=n_IJ_IM_z \, \sigma_z\tau_\alpha
\]
and a suitable continuation to complex energy \(E\) is assumed. The explicit form of the mean field Green function can be written as,
\begin{equation}
  \label{e:g}
  g(E,\bm k) = \begin{pmatrix}
    g_{\uparrow} (E,\bm k) & 0 \\
  0 & g_{\downarrow} (E,\bm k)
  \end{pmatrix}
\end{equation}
where
\begin{align}
  \label{e:gup}
  g_\uparrow &= \frac{1}{\Delta_\uparrow} \left[
    (E + \mu_\uparrow + Dk^2)\tau_0 +
  (m_\uparrow - Bk^2)\tau_z + \bm k \cdot \bm \tau \right] \\
  \label{e:gdo}
  g_\downarrow &= \frac{1}{\Delta_\downarrow} \left[
    (E + \mu_\downarrow + Dk^2)\tau_0 +
  (m_\downarrow - Bk^2)\tau_z - \bm k \cdot \bm \tau^* \right]
\end{align}
where we defined spin dependent chemical potential (\(\mu_{\uparrow\downarrow}\)) and mass (\(m_{\uparrow\downarrow}\)) as,
\begin{equation}
  \label{e:muF}
  \mu_\uparrow = \mu - v\,, \quad \mu_\downarrow = \mu + v\,, \quad
  m_\uparrow = m_\downarrow = m
\end{equation}
with \(v=n_IJ_IM_z\), for the ferromagnetic case (note that the noise term \(v\) enters through the chemical potential), and
\begin{equation}
  \label{e:muA}
  \mu_\uparrow = \mu_\downarrow = \mu\,, \quad
  m_\uparrow = m + v\,, \quad m_\downarrow = m - v
\end{equation}
for the antiferromagnetic case (here the random potential adds to the mass term); the determinants \(\Delta_{\uparrow,\downarrow} = \Delta_{\uparrow,\downarrow}(E,k)\), are given by,
\[
  \Delta_{\uparrow,\downarrow} = (E+\mu_{\uparrow,\downarrow} - k^2)^2 -
  (m_{\uparrow,\downarrow} - Bk^2)^2 - k^2
\]
(note that they only depend on the wavenumber modulus).

The random averaged Green function is defined in terms of the self-energy,
\begin{equation}
  \label{e:G}
  G(E,\bm k) = \langle (E - H)^{-1} \rangle
             = [g^{-1}(E,\bm k)-\Sigma(E)]^{-1}\,.
\end{equation}
where the self-energy \(\Sigma(E)\), is computed from a perturbation series in powers of the random potential (the dependence on the kind of magnetic order is taken in the form of the parameters (\ref{e:muF}) and (\ref{e:muA})). It depends only on the energy: after averaging over the impurities postions and magnetic moments orientations (isotropic in the plane), one should recover translation invariance. To second order,\cite{Bruus-2004xy}
\begin{equation}
  \label{e:Sigma}
  \Sigma(E)= \sum_{\bm q}
  \langle \Delta V(\bm q - \bm k) g(\bm q) \Delta V(\bm k - \bm q) \rangle 
\end{equation}
where \(\Delta V= V - \langle V\rangle\). Only diagonal terms survive to the disorder averaging and integration over the polar angle of the internal wavevector. Transforming the sum into an integral, and performing the matrix multiplication and noise averaging, one gets,
\begin{equation}
  \label{e:Sa}
  \Sigma(E) = n_IJ_I^2 \int_0^{k_*} 
  \frac{qdq}{2\pi} \left[ v_z g_z(E,q) + v_\perp g_\perp(E,q) \right]
\end{equation}
where we introduced a lattice cutoff \(k_*=\pi\) (at the border of the Brillouin zone), and we defined the diagonal matrices,
\begin{multline}
  \label{e:gz}
  g_z = \mathrm{diag}\left[
    \frac{E + \mu_\uparrow + Dk^2}{\Delta_\uparrow}\tau_0 + \frac{m_\uparrow - Bk^2}{\Delta_\uparrow}\tau_z, 
  \right.\\
  \left.
    \frac{E + \mu_\downarrow + Dk^2}{\Delta_\downarrow}\tau_0 + \frac{m_\downarrow - Bk^2}{\Delta_\downarrow}\tau_z
  \right]
\end{multline}
and
\begin{equation}
  \label{e:gperp}
  g_\perp = \sigma_0\tau_x g_z \sigma_0 \tau_x
\end{equation}
the same matrix as \(g_z\) with the spin sectors exchanged, and
\begin{align}
  v_z &= \langle S_z^2 \rangle - n_I^2M_z^2 = (1-n_I^2)M_z^2 + \frac{1}{3}(1-M_z)^2\,, \nonumber \\
  v_\perp &= 1 - \langle S_z^2 \rangle = \frac{2}{3}(1-M_z)(1+2M_z)\,,
  \label{e:vzvp}
\end{align}
in the fully polarized case \(M_z=1\), only the first term in \(v_z\) remains; the fluctuation effects are maximized for a impurity concentration \(n_I=1/\sqrt{3}\). In fact, the \(q\) integral in (\ref{e:Sa}) is, in the continuous limit, logarithmically divergent; hence, keeping only the dominant terms contributing to the real part near the Fermi energy \(E_F\), one obtains,
\begin{multline}
  \label{e:reS}
  \mathrm{Re}\,\Sigma_{\uparrow\downarrow} = \frac{n_IJ_I^2}{8\pi}
    \frac{B\tau_z-D \tau_0}{B^2-D^2} 
    \left[
    v_z \log\left| 
      \frac{(B^2-D^2)k_*^4}{\Delta_{\uparrow\downarrow}(E_F,0)} 
    \right| \right. \\
    + \left.
    v_\perp \log\left| 
      \frac{(B^2-D^2)k_*^4}{\Delta_{\downarrow\uparrow}(E_F,0)}
    \right|
    \right] 
\end{multline}
where the dependence on the impurities polarization is through the parameters \(\mu_{\uparrow\downarrow}\) and \(m_{\uparrow\downarrow}\) present in the determinants \(\Delta\). The term proportional to \(\tau_0\) renormalizes the Fermi level (\(\mu\)) and the one in \(\tau_z\) renormalizes the gap (\(m\)). We observe that longitudinal (\(v_z\)), and transverse (\(v_\perp\)), fluctuations introduce corrections of the gap and energy levels that depend on spin; in particular, transverse fluctuations couple the two spin polarizations, through which flipping of the spin by scattering becomes possible.

This renormalization also has an impact on the behavior of the spin dependent edge states. A simple computation\cite{Zhou-2008fk} leads to the following conditions for the existence of edge states,
\begin{multline}
  \label{e:edgef}
  \frac{m_{\uparrow\downarrow} }{B}=\frac{m}{B} 
  +
  \frac{n_I J_I^2}{8\pi(B^2-D)^2} \left[ v_z
    \log \left|
      \frac{(B^2-D^2)k_*^4}{(E_F \mp v)^2 - m^2}
    \right| \right. \\
    + \left. v_\perp \log \left|
        \frac{(B^2-D^2)k_*^4}{(E_F \pm v)^2 - m^2}
    \right| \right] > 0
\end{multline}
(\(v=n_IJ_IM_z\), the upper and lower signs correspond to spin up and spin down, respectively) for the ferromagnetic case, and
\begin{multline}
  \label{e:edgea}
  \frac{m_{\uparrow\downarrow} }{B}=\frac{m \pm v}{B} 
  +
  \frac{n_I J_I^2}{8\pi(B^2-D^2)} \left[ v_z \log
  \left|
    \frac{(B^2-D^2)k_*^4}{E_F^2 - (m \pm v)^2} \right| \right. \\
    + \left. v_\perp \log
  \left|
    \frac{(B^2-D^2)k_*^4}{E_F^2 - (m \mp v)^2}  \right| 
  \right]> 0
\end{multline}
for the antiferromagnetic case. In the ferromagnetic case the main effect is a decrease of the characteristic penetration length and opposite shifts of the Fermi energy depending on spin. In the antiferromagnetic case, the characteristic length of the spin down state tends to increase, making this channel penetrate into the bulk material.\cite{Liu-2008ys,Li-2013qf} This effect may be considered as the onset of the \(G=2\) to \(G=1\) transition.

In principle, the logarithmic term in (\ref{e:reS}) can change sign for strong enough disorder, typically when \(v \gg B\), inducing a qualitative change in the topological properties of the insulator. Although this regime is somewhat outside the validity of the perturbation expansion, it indicates that the effect of fluctuations can be nontrivial, as in the topological insulator transition driven by Anderson localization.\cite{Li-2009kx}

\subsection{Phase diagram}

\begin{figure*}
  \centering
  \includegraphics[width=0.49\textwidth]{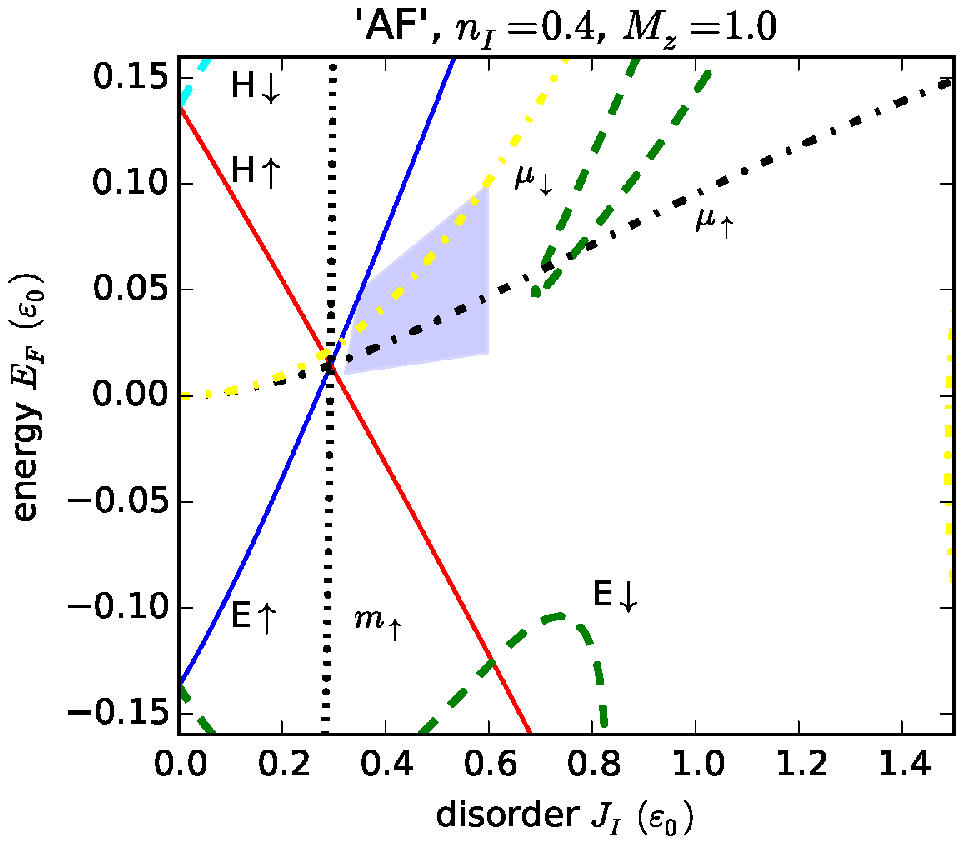}\hfill%
  \includegraphics[width=0.49\textwidth]{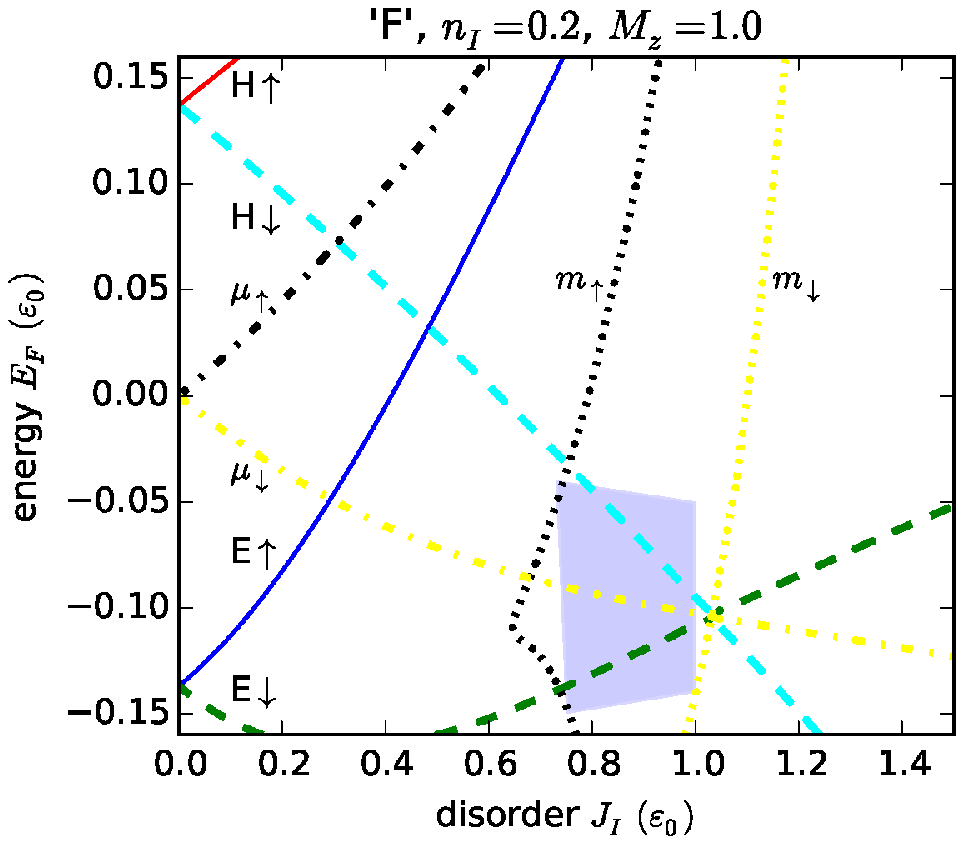}
  \caption{\label{f:ti4} Phase diagram computed from the self-consistent Born approxiamtion. Antiferromagnetic (left) and ferromagnetic (right) full polarized cases. The blue shaded region corresponds schematically to the $G=1$ phase. Solid lines stand for spin up bands, dashed lines for spin down bands; dotted lines separate the topological phase (on their left side) to the trivial ones (towards strong disorder); dot-dashed lines are for the chemical potential levels. }
\end{figure*}

The perturbation expansion, even within the mean field approximation that accounts for the spin splitting, cannot be extended to finite disorder. Moreover, non-dominant terms neglected in the calculation of the integral in (\ref{e:Sa}), possess symmetry properties different from the kept ones, and their contribution can become significant at finite disorder. It is therefore convenient to compute the self-energy using instead the self-consistent approximation by replacing \(g\) in (\ref{e:Sigma}) by the full averaged Green function (\ref{e:G}); this allows a resummation of all noncrossing diagrams and extend the range of validity of the renormalization of the bare parameters as a function of the disorder.

The numerical resolution of the implicit equation for \(\Sigma\) gives the renormalized values of the up and down masses and chemical potentials (barred quantities):
\begin{multline}
  \mathrm{Re}\,\Sigma \equiv 
  -\mathrm{diag}[ (\bar{m}_{\uparrow}-m_\uparrow) \tau_z +
    (\bar{\mu}_{\uparrow}-\mu_\uparrow)\tau_0, \\
    (\bar{m}_{\downarrow}-m_\downarrow) \tau_z - 
    (\bar{\mu}_{\downarrow}-\mu_\downarrow)\tau_0 ]
  \end{multline}
as a function of the Fermi energy \(E_F\) and disorder strenght \(J_I\). Explicitly, the self-consistent equations for the renormalized parameters are:
\begin{align}
  \begin{split}
    \bar{\mu}_\uparrow &= \mu_\uparrow -
    \frac{n_IJ_I^2}{N^2} \sum_{k_x,k_y}
    \bigg[
    \frac{v_z}{\Delta_\uparrow} (E_F + \bar{\mu}_\uparrow + D_k)
    \\ & \qquad +
    \frac{v_\perp}{\Delta_\downarrow} (E_F + \bar{\mu}_\downarrow + D_k)
    \bigg]
  \end{split}
\\
  \begin{split}
    \bar{\mu}_\downarrow &= \mu_\downarrow -
    \frac{n_IJ_I^2}{N^2} \sum_{k_x,k_y}
    \bigg[
    \frac{v_z}{\Delta_\downarrow} (E_F + \bar{\mu}_\downarrow + D_k)
    \\ & \qquad +
    \frac{v_\perp}{\Delta_\uparrow} (E_F + \bar{\mu}_\uparrow + D_k)
    \bigg]
  \end{split}
  \label{e:renor}
\end{align}
 and
\begin{align}
  \bar{m}_\uparrow &= m_\uparrow - \frac{n_IJ_I^2}{N^2} \sum_{k_x,k_y}
    \bigg[
      \frac{v_z}{\Delta_\uparrow}(\bar{m}_\uparrow - B_k)+
      \frac{v_\perp}{\Delta_\downarrow}(\bar{m}_\downarrow - B_k)
    \bigg]\\
    \bar{m}_\downarrow &= m_\downarrow - \frac{n_IJ_I^2}{N^2} \sum_{k_x,k_y}
    \bigg[
      \frac{v_z}{\Delta_\downarrow}(\bar{m}_\downarrow - B_k)+
      \frac{v_\perp}{\Delta_\uparrow}(\bar{m}_\uparrow - B_k)
    \bigg]
\end{align}
where \(B_k = B(4-2 \cos k_x -2 \cos k_y)\), \(D_k = D(4-2 \cos k_x -2 \cos k_y)\), and
\[
  \Delta_\uparrow =  (E_F + \bar{\mu}_\uparrow + D_k)^2-
     (\bar{m}_\uparrow - B_k)^2 - \sin^2k_x-\sin^2k_y
\]
and a similar expression for the spin down determinant \(\Delta_\downarrow\). We note that fluctuations introduce a coupling between spin up and spin down parameters that vanish in the full polarized case (\(v_\perp=0\)). Indeed, as already present in Eqs.~\ref{e:reS} of the perturbation series, in-plane fluctuations (proportional to \(v_\perp\)), break \(\sigma_z\)-spin conservation, and couple spin orientations through scattering off the impurities.

We represented in Fig.~\ref{f:ti4}, the gap edges \(\mu_{\uparrow\downarrow} = \pm m_{\uparrow\downarrow}\), for the antiferromagnetic (left) and ferromagnetic (right) cases, of the electron (E$\uparrow$, E$\downarrow$) and hole (H$\uparrow$, H$\downarrow$) bands, together with the band inversion thresholds for the spin up (\(m_\uparrow=0\)) and down (\(m_\downarrow=0\)) subbands, and the spin dependent chemical potentials (\(\mu_\uparrow,\mu_\downarrow\)). 

In the antiferromagnetic case, the crossing of H$\uparrow$ and E$\uparrow$, coinciding with the \(m_\uparrow=0\) threshold, points out the opening of a gap, and the restoring of a topologically trivial state of the spin up subband. The \(G=2\) and \(G=1\) regions should locate on the left and right of the \(m_\uparrow=0\) vertical line, respectively (as schematically represented by the blue shaded region, around the \(\mu_\downarrow\) line).

In the ferromagnetic case, a remarkable effect arises: the bending towards positive energies of both E$\uparrow$ and E$\downarrow$ subbands. Due to this bending the electron bands can cross the corresponding hole bands with the same spin. The \(m_\uparrow\) line intersects the H$\uparrow$ and E$\uparrow$ crossing, as in the antiferromagnetic case, and a \(G=1\) region can develop on its higher disorder strength side. This region is delimited by the crossing of the spin down subbands and the \(m_\downarrow=0\) line.

These results suggest that, although the mechanisms are different, the effective band structure leading to the anomalous quantum Hall state is essentially the same for the two coupling modes (equal or opposite spin splitting sign of the electron and hole subbands). In the ferromagnetic case, it is a strong disorder effect.

\subsection{Conclusion}

The main result of the present study is that a magnetically doped two-dimensional topological insulator supports different quantized transport regimes as a function of the disorder strength. The transition between a topological insulator phase, characterized by the presence of two spin polarized channels, to a Chern insulator phase, with one spin conduction channel, is related to the selective suppression of one of the spin states. In particular, we observed the emergence of this anomalous quantum Hall state, in the case of equal sign coupling of the electron and hole subbands with the magnetic impurities (ferromagnetic case). This is a disorder driven effect that situates beyond the range of validity of the mean field approximation. The complete neglect of fluctuations, leads to a simple renormalization of the chemical potential, without incidence on the mass gap. At variance, in the antiferromagnetic case (opposite signs couplings for electrons and holes), this approximation is enough to explain the opening of a gap for one spin band. 

More generally, the effect of disorder manifests by a renormalization of the mass and chemical potentials, which become spin dependent due to the fluctuations contributions. In the ferromagnetic case, strong disorder causes a bending of the bands that allows a nontrivial crossing. As revealed by the local density of states and the behavior of the local currents, the appearance in this case of the Chern insulator state, is related to the localization of one of the spin bands. As a consequence, the device behaves as a spin filter, with accumulation of the allowed spin conduction band on the opposite lead.

\begin{acknowledgments}
We thank Steffen Schäfer, Roland Hayn and Tineke van den Berg for useful discussions. LR and AV acknowledge Thierry Martin and the Centre de physique théorique (Marseille), where part of this work was completed. Numerical calculations were performed at the Mésocentre and Plateforme Technique de Calcul (Marseille).
\end{acknowledgments}

\end{document}